
\input harvmac.tex
%
{\nopagenumbers
\rightline{SPhT/92-159}
\rightline{hep-th/9212106}
 \vbox{
  \vphantom{0}\vskip.5truein
  \centerline{\bf NON-PERTURBATIVE EFFECTS IN MATRIX MODELS}
  \centerline{\bf AND VACUA OF TWO DIMENSIONAL GRAVITY}
 }
\vskip.65truein
\centerline{Fran\c cois DAVID
\footnote{$^\dagger$}{Physique Th\'eorique CNRS}}
\vskip 2.ex
\centerline{\it Service de Physique Th\'eorique
\footnote{$^\star$}{Laboratoire de la Direction des Sciences de la Mati\`ere
du Commissariat \`a l'Energie Atomique}}
\centerline{\it CE Saclay, 91191 Gif/Yvette CEDEX, FRANCE}
\vskip .5ex
\vskip .65truein
\baselineskip 14pt plus 1pt minus 1pt

\centerline{\bf Abstract}
\bigskip
The most general large $N$ eigenvalues distribution for the one matrix model
is shown to consist of tree-like structures in the complex plane.
For the $m=2$ critical point, such a solution describes the strong coupling
phase of $2d$ quantum gravity ($c=0$ non-critical string).
It is obtained by taking combinations of complex contours in the matrix
integral, and the relative weight of the contours is identified with the
non-perturbative ``$\theta$-parameter" that fixes uniquely the solution of the
string equation (Painlev\'e I).
This allows to recover by instanton methods results on the non-perturbative
effects obtained by the Isomonodromic Deformation Method, and to construct for
each $\theta$-vacuum the observables (the loop correlation functions) which
satisfy the loop equations.
The breakdown of analyticity of the large $N$ solution
is related to the existence of poles for the loop operators.
\vfill
\leftline{December 1992}
\eject
}
\baselineskip 14pt plus 1pt minus 1pt
\pageno=1

The discovery of the ``double scaling solutions" of the matrix models
\ref\BreKaz{E. Br\'ezin and V. A. Kazakov, Phys. Lett. 236B (1990) 144.}
\ref\DouShe{M. R. Douglas and S. H. Shenker, Nucl. Phys. B 335 (1990) 635.}
\ref\GroMig{D. J. Gross and A. A. Migdal, Phys. Rev. Lett. 64 (1990) 127.}
led to important progress in the understanding of string theories
in $d\le 2$ backgrounds and of $2d$ gravity (see
\ref\Cargese{{\sl ``Random Surfaces, Quantum Gravity and Strings"},
NATO ASI Series B: Physics Vol. 262, O. Alvarez, E. Marinari and P. Windey
Eds. , Plenum Press, New York (1991).}
\ref\Jerus{{\sl `` Two dimensional quantum gravity and random surfaces"},
Jerusalem Winter School for Theoretical Physics Vol. 8,
D. J. Gross, T. Piran and S. Weinberg Eds.,
World Scientific, Singapore (1991).}
for reviews).
However, the important issue of the non-perturbative status of some of these
theories remains unclear, in particular for $2d$ gravity coupled to unitary
matter for $c\le 1$.
In this letter, we discuss some of these questions in the framework of
the Hermitian one matrix models.
We shall show that a simple generalization of the complex integration contour
prescription
\ref\loopmoi{F. David, Mod. Phys. Lett. A5 (1990) 1019.}
\ref\SilYel{P. G. Silvestrov and A. S. Yelkovsky, Phys. Lett. 251B (1990) 525.}
, which allows to construct non-perturbative --- but in general complex ---
solutions of the string equations and of the continuous loop equations,
leads to real non-perturbative solutions of these equations.
This generalization, which consists in taking combinations of inequivalent
integration contours, has been already discussed by Fokas, Its and Kitaev
\ref\FokItsKit{A. S. Fokas, A. R. Its and A. V. Kitaev, Commun. Math. Phys.
142 (1991) 313, and 147 (1992) 395.}
in the framework of the Isomonodromic Deformation Method (IDM) approach to
the string equations
\ref\moore{G. Moore, Commun. Math. Phys. 133 (1990) 261.}
, but does not seem to have attracted much attention.
Our treatment is based on the BIPZ solution of the one matrix model
\ref\BIPZ{E. Br\'ezin, C. Itzykson, G. Parisi and J.-B. Zuber,
Commun. Math. Phys. 59 (1978) 35.},
and follows our previous analysis of
\ref\Davpha{F. David, Nucl. Phys. B 348 (1991) 507.}.
We shall show that in the limit $N\to\infty$, new solutions for the eigenvalues
(e.v.) distribution exist, which have not been discussed before.
They correspond to a distribution of e.v. along tree-like structures in the
complex plane.
Moreover, these solutions depend non-analytically of the coupling constant
of the matrix model, and will be associated with the sectors with an infinite
number of poles of the string equation solutions.
The non-perturbative parameter which characterizes the non-perturbative
solutions is simply related to the different weights chosen for the contours,
and our treatment allows to recover easily by instanton methods some results
of \SilYel\FokItsKit .
In addition, we show that to each real solution of the string equation
is associated a prescription for the asymptotics of the loop operators
which defines uniquely observables (i.e. macroscopic loop v.e.v.) which
obey the loop equations.
Finally we shall show that these new solutions allow to explain the properties
of the solutions for the double well matrix models recently discussed
by Brower, Deo, Jain and Tan
\ref\BrDeJaTa{R. C. Brower, N. Deo, S. Jain and C. I. Tan,
{\it preprint 1992}.}.
\medskip
In the matrix model formulation of $2d$ gravity, the partition function $F$
(sum over orientable connected 2-dimensional Riemannian spaces)
is discretized into a sum over triangulations,
and is written as the logarithm of the partition function $Z$ for the Hermitian
one-matrix model ($F=\ln Z$), which after diagonalization of the matrix $\Phi$
can be written as an e.v. integral
\eqn\evInt{
Z_N\ =\ {\bf C}_N\,\int \,\prod_{i=1}^N \, d\lambda_i {\rm
e}^{-N\,V(\lambda_i)}
\ \Delta_{\scriptscriptstyle N}(\lambda_i)^2\qquad;\qquad
\Delta_{\scriptscriptstyle N}(\lambda_i)\ =\
\prod_{i<j} (\lambda_i-\lambda_j)
}
where ${\bf C}_N$ is a normalization factor, $\Delta_{\scriptscriptstyle N}$
the Vandermonde determinant and $V$ the matrix potential.
The integral \evInt\ can be calculated in terms of the matrix elements of
the operator $Q\ :\ \pi_n(\lambda)\mapsto \lambda\pi_n(\lambda)$, where the
$\pi_n$ are orthonormal polynomials for the measure
$d\lambda{\rm e}^{-NV(\lambda)}$.
In the double scaling limit, $N\to\infty$ and $V\to V_{\rm critical}$ while
$x=1-{n \over N}$ becomes a continuous parameter.
Then $Q$ becomes a second order differential operator of the form
\eqn\Qcont{
Q\ =\ -{d^2\over dx^2}\,+\,2u(x)
}
$u$ is the string susceptibility
\eqn\strsusc{
u(t)\ =\ -{\partial^2 F\over \partial t^2}
}
where $t$ is the renormalized cosmological constant.
For the $m=2$ critical point (pure $c=0$ gravity), $t$ scales with $N$ as
$t\sim N^{4/5}$, and $u$ satisfies the Painlev\'e I string equation
\eqn\Painleve{
-{1\over 6}\,{\partial^2 u \over \partial t^2}\,+\,u^2\ =\ t\qquad;\qquad
u\sim\sqrt{t}\qquad t\to +\infty
}
It is known that \Painleve\ fixes uniquely the terms of the asymptotic
expansion of $u$ (in powers of $t^{(1-5k)/2}$) as $t\to +\infty$, but that
the corresponding series is not Borel summable, and that the solutions of
\Painleve\ form a one parameter family of ``simply truncated solutions"
\ref\Boutroux{P. Boutroux, Ann. Ecole Norm. 30 (1913) 265, and 31 (1914) 99.},
which differ by exponentially small
terms of the form
\eqn\npterm{
\delta u\ \propto\ t^{-1/8}\,{\rm e}^{-{4\over 5}2 \sqrt{3} t^{5/4}}
}
\def\II{\hbox{I\hskip-0.1em I}}
\def\III{\hbox{I\hskip-0.1em I\hskip-.1em I}}
\def\IV{\hbox{I\hskip-0.1em V}}
The real solutions of \Painleve\ have an infinite series of double poles
(with residues $1$) on the negative real axis.
If one divides the complex plane into 5 sectors ${\bf s}={\rm I},\ldots,{\rm
V}$
(which correspond to
$({\bf s}-1)\,2\pi/5 < {\rm Arg} (t) < {\bf s}\,2\pi/5$),
these poles extend to an
infinite network of poles in the sectors \II, \III\ and \IV,
so that the asymptotics $u\sim\sqrt{t}$ holds only in the two pole-free sectors
I and V.
\medskip
It was suggested in \loopmoi ,
and shown more precisely in \Davpha\SilYel\FokItsKit , that, if one
constructs the $m=2$ theory by starting from a cubic potential of the form
\eqn\CubPot{
V(\lambda)\ =\ -\lambda^3\,+\,\ldots
}
and defines the integral \evInt\ by choosing as integration contour for the
$\lambda_i$'s the complex contour ${\cal C}_+$ (resp. ${\cal C}_-$)
which goes from $-\infty$ to $j\infty$ ($j={\rm e}^{i\pi/3}$)
(resp. ${\bar j}\infty$), one obtains the ``simply truncated solution"
$u_+$ (resp. $u_-$) of \Painleve\ which has poles only in the sector \II\
(resp. \IV) and satisfies the asymptotics $u\sim \sqrt{t}$ in the remaining
four pole-free sectors .
 From these two solutions, which are analytic on the real axis, one can
construct without ambiguity the operator $Q$ \Qcont , which is not Hermitian
anymore, and the loop operators $w(p)$
\ref\BDSS{T. Banks, M. R. Douglas, N. Seiberg and S. H. Shenker,
Phys. Lett. 238B (1990) 279.}
(where $p$ is the loop momentum), which satisfy the loop equations
\loopmoi
\ref\DVV{
R. Dijkgraaf, E. Verlinde and H. Verlinde, Nucl. Phys. B348 (1991) 435.}
\ref\FKN{
M. Fukuma, H. Kawai and R. Nakamaya, Int. J. Mod. Phys. A6 (1991) 1385.}
{}.
\nfig\fInt{The three contours for the cubic potential}
\medskip
In fact a straightforward generalization of this prescription is to
consider linear combinations of the two contours, i.e. to replace
\eqn\ContComb{
\int_{{\cal C}_\pm}\! d\lambda\ \longrightarrow \ c_+\,\int_{{\cal
C}_+}d\lambda\,+\,
c_-\,\int_{{\cal C}_-}d\lambda}
Indeed, the partition function $Z$ will still be real (for real $V$) if the
weight $c_\pm$ are complex conjugate
\eqn\RealW{
c_\pm \ =\ {1\over 2}\pm i\theta}
With this prescription, the orthogonal polynomial method still works, and the
recurrence relations (discrete string equations) still hold.
Therefore, if the double scaling limit exists, one should still obtain some
solution of the string equation \Painleve .

As already mentioned, it has been shown in \FokItsKit , within the IDM
approach, that this is indeed the case, and that there is a one to one
correspondence between the weight ratio $c_+/c_-$ and the simply truncated
solution of \Painleve\ which is obtained in the double scaling limit.
To recover this result in the BIPZ approach, let us consider the matrix
model integral \evInt\ in the large $N$ limit.
The eigenvalue probability density
${1\over N}\sum\limits_{i=1}^N \delta(\lambda-\lambda_i)$ becomes the classical
density $v(\lambda)$.
It is convenient to consider the function
\eqn\Ffunct{
F(\lambda)\ =\ \int d\mu\,{v(\mu)\over \lambda-\mu}\ =\ \lim_{N\to\infty}
{1\over N}\langle{\rm Tr}\left({1\over \lambda-\Phi}\right)\rangle
\qquad .}
It can be shown, from the saddle point equations for the effective
action for $v$, or through the loop equations \loopmoi , that $F$ must be of
the form
\eqn\Fgen{
F(\lambda)\ =\ {1\over 2}\left[ V'(\lambda)+\sqrt{Q(\lambda)}\right]\qquad;
\qquad Q(\lambda)\ =\ V'(\lambda)^2+4N(\lambda)}
where $N(\lambda)$ is some polynomial of degree
\eqn\degree{
{\rm degree}\,N\ =\ {\rm degree}\, V\,-\,2\,=\,m-1}
Generically, $Q$ has $2m$ complex zeros, which correspond to square root cuts
for $F$.
 From \Ffunct\ the e.v. density $v$ is proportional to the discontinuity of $F$
along the cuts, and can be reconstructed from $F$.
The normalization $\int d\lambda\, v=1$ implies that $F\sim\lambda^{-1}$ at
$\infty$ and this fixes the coefficient of the leading term of $N$.
Requiring that $F$ has only
one cut (as done for instance in \BIPZ ) implies that $Q$ has $m-1$ double
zeros and this fixes uniquely $N$.
However, in general $F$ may have several cuts.

Let us label by $\alpha$ the cuts and by $x_\alpha$ the fraction of e.v. along
each cut ($x_\alpha\ge 0$ and $\sum x_\alpha =1$).
The e.v. density $v$ must minimize the action
\eqn\effact{
S\,=\,\int d\lambda\,v(\lambda)V(\lambda)\,-\,
\int\int d\lambda\,d \mu\,v(\lambda)\,v(\mu)\,\ln |\lambda-\mu|\,+\,
\sum_\alpha \Gamma_\alpha\left(x_\alpha-\int_\alpha d\lambda v(\lambda)\right)
}
where $\Gamma_\alpha$ are Lagrange multipliers.
In fact \Fgen\ is the most general solution when one extremizes
$S$ w.r.t. variations of the density  which do not change the
$x_\alpha$'s. As a consequence, the effective potential $\Gamma(\lambda)$ for
one e.v. in the background created by the $N-1$ other e.v.'s
\eqn\effpot{
\Gamma(\lambda)\ =\ V(\lambda)\,-\,2\,\int\hskip-1.em- d\mu\,v(\mu)\,
\ln(\lambda-\mu)
}
is constant along each cut $\alpha$, and equal to $\Gamma_\alpha$
\eqn\Gcst{
\Gamma(\lambda)\ =\ \Gamma_\alpha\qquad{\rm if}\qquad\lambda\in\alpha
}
so that the total action is
\eqn\totact{
S\ =\ {1\over 2}\,\sum_\alpha\,\left[ \int_\alpha d\lambda\,v(\lambda)
V(\lambda)\,+\,x_\alpha\Gamma_\alpha \right]
}
The remaining constraints which fix $N$ are the following:
\item{(a)} The e.v. fractions $x_\alpha$ must be real\par
\item{(b)}
One must minimize ${\rm Re}(S)$ w.r.t. variations of the $x_\alpha$'s,
subjected to the constraints that $x_\alpha \ge 0$ and that
$\sum x_\alpha =1$.
Since from \effact\ ${\partial S\over \partial x_\alpha}=\Gamma_\alpha$,
this implies that the real part of the effective potential is the same along
all
the cuts:\par
\eqn\GamEqu{
{\rm Re}(\Gamma_\alpha)\ =\ \Gamma_0\qquad\hbox{unless}\
x_\alpha\ =\ 0 }
\medskip
\nfig\fCont{The contours in \Constraint\ corresponding to
constraints (a) and (b)}
In fact the constraints (a) and (b) can be recast in the same form:
\eqn\Constraint{
{\rm Im}({\cal I}_a)\ =\ 0\qquad,\qquad {\cal I}_a\,=\,\oint_{{\cal C}_a}
{d\lambda\over 2i\pi}\,F(\lambda)}
where ${\cal C}_a$ is any contour encircling a pair of zeros of $Q$.
Indeed, if ${\cal C}$ encircles a cut $\alpha$, ${\cal I}=x_\alpha$,
while if ${\cal C}$ encircles the end points of two different cuts $\alpha$
and $\beta$, it follows from the fact that away from the cuts,
$\Gamma'(\lambda)=V'(\lambda)-2F(\lambda)$ (obtained by taking the derivative
of \effpot ), that ${\cal I}={1\over i\pi}(\Gamma_\beta-\Gamma_\alpha)$
(see \fCont ).
Since there are $2m-2$ independent contours, \Constraint\ fixes the $m-1$
remaining coefficients of $N$.
Let us note that if $Q$ has a double zero, two independent constraints
\Constraint\
are automatically satisfied, since for any contour ${\cal C}$ which encircles
the double zero and at most one single zero, ${\cal I}=0$. Therefore we expect
in general to have 1 solution of \Constraint\ with no double zeroes,
$2m-1$ solutions with one double zero, etc$\ldots$
Some of these different solutions will be excluded because:
\item{(c)} Some $x_\alpha$'s are $<0$.
\item{(d)} The contours of integration for the e.v. cannot pass through one of
the cuts $\alpha$.\par
\noindent Finally, among the remaining solutions, it is the one with minimal
${\rm Re}(S)$ (real part of the action \effact ) which is the physical saddle
point.
\medskip

\nfig\fCut{Schematic picture of
the e.v. distribution (black line) and the ${\rm Re}(\Gamma)>0$ domains
(grey) in the $p$ complex plane for the generic solution of the $m=2$
critical point:
(a) real $t>0$ (and sectors I and V); (b) real $t<0$ (and sector \III );
(c) $t$ in sector \II ; (d) $t$ in sector \IV .
}
Let us discuss explicitly the case of the $m=2$ critical point.
We start from the potential $V(\lambda)=-\lambda^3/3+g\lambda$.
In the critical regime we rescale
$g_c-g\simeq a^2 t$, $\lambda-\lambda_c\simeq a p$ \Davpha .
$t$ is the renormalized cosmological constant and $p$ the loop momentum.
The scaling parameter (short-distance cut-off) $a$ is defined so that the
double
scaling limit is obtained by taking $N\to\infty$,
$g_{\rm string}^{-2}=N^2a^5=1$.
In the planar scaling limit ($N=\infty$, then $a\to 0$), the general solution
\Fgen\ for $F(\lambda)$ becomes
\eqn\Fscal{
F(\lambda)\ \to\ w(p)\,=\,a^{5/2}\,{2\over 3}\sqrt{p^3-3tp+{\bf c}}
}
where ${\bf c}$ is the only parameter to be determined in the polynomial
$N(\lambda)$ which is relevant in the scaling limit. It corresponds to
the v.e.v of the puncture operator \loopmoi .
In the weak coupling phase $t>0$, the saddle point is the standard one cut
solution \BIPZ\loopmoi:
\eqn\solpert{
{\bf c}\,=\,2\,t^{3/2}\qquad;\qquad
w(p)\,=\,a^{5/2}\,{2\over 3}\,(t^{1/2}-p)\,\sqrt{p+2\,t^{1/2}}
}
The e.v. are located on the real axis along the cut $]-\infty,-2t^{1/2}]$
(see {\fCut\ (a)}).
The action for this solution scales as
\eqn\actpert{
S\ \propto\ a^{5/2}\,t^{5/2}
}
There are other unphysical solutions which violate (c) or (d).

This solution can be analytically continued into the strong coupling
unstable phase $t<0$. The two complex conjugate solutions describe e.v.'s
still located along a single arc \Davpha , and correspond to the
triply truncated solutions of \Painleve .
 From \actpert\ they have a purely imaginary action.
However for $t<0$ the constraints \Constraint\ have another solution, where
$w(p)$ has now three branch points.
For this solution {\bf c} is given by
\eqn\solnpert{
{\bf c}\,=\,c\,(-t)^{3/2} \qquad;\qquad c>0
}
and it corresponds to the branched distribution of the e.v's depicted on
{\fCut\ (b)} .
The density of e.v. along the negative real axis vanishes at
the real branch point $p_0$ as $\sqrt{p_0-p}$, but there are two arcs
starting from $p_0$ toward the two other complex conjugate branch points
$p_\pm$.
Therefore an equal fraction $x_+=x_-$ of e.v. are sitting along these two arcs.
Moreover the action for this solution is real and negative, and
therefore it is generically the dominant one.
Away from the negative real axis, the branched solution still exists,
but with asymmetric branches ($x_+\ne x_-$), provided that one stays in the
sector \III\ ($-\pi/5 <{\rm Arg}(-t) < \pi/5$),
and still has a lower action than the perturbative one.
Moreover, since the constraints \Constraint\ are non-analytic, this solution
does not depend analytically of $t$, (in other word, the number $c$ in
\solnpert\ depends on ${\rm Arg}(t)$).

Finally, in the sectors \II\ and \IV , another kind of solutions, with
two cuts, is dominant (see {\fCut\ (c--d)}).
These solutions are still non-analytic in $t$.
In the sectors I and V, the perturbative analytic solution \solpert\ is
the physical one.

As is clear from \fCut , for the general integration prescription \ContComb ,
one can obtain these new, non-analytic, solutions in the
sectors \II, \III\ and \IV.
Since they have lower action than the analytic solution, they will dominate
the large $N$ limit, unless $c_+$ or $c_-$ is zero.
\medskip

This new solution for the e.v. distribution is associated to the
simply truncated solutions of \Painleve . This can be seen as follows.
First, we have seen that these solutions are non analytic in $t$ in the
sectors where simply truncated solutions have poles.
Since double poles of $u(t)$ should correspond to simple zeros of the $\tau$
function, which corresponds to the partition function $Z$ for
the matrix model, and since in the large $N$ limit $Z$ is obtained from
the action $S$ for the saddle point e.v. configuration by $Z=\exp (N^2 S)$,
using Cauchy formula for the derivative of $\ln(Z)$ and Stoke's formula we
obtain the estimate for the number of zeros in a domain $D$
\eqn\zerdens{
{\#\ {\hbox{of zeros}}\atop{\hbox{in domain $D$}}}\ =\
\oint_{\partial D}
{d g\over 2i\pi}\,
{\partial\ln(Z)\over\partial g}
\ \simeq\ N^2\,\int\hskip-.7em\int_D {dg\,d{\bar g}\over
4\pi}\,{\partial\over\partial g}\,
{\partial\over\partial{\bar g}}\,S(g,{\bar g})
}
Thus the new solution describes a partition function with a positive density
of zeros $\rho\propto\partial{\bar\partial}S$
in the three sectors \II, \III, \IV, which scales as
$\rho\propto N^2 a^{5/2} |t|^{1/2}f(\phi)$, where $\phi={\rm Arg}(t)$.

One can make the identification more precise by relating the constraints
\Constraint\ to the asymptotics of the simply truncated solutions of \Painleve
{}.
Following \Boutroux\ (see
\ref\Hille{E. Hille, {\sl ``Ordinary Differential Equations in the Complex
Domain"}, Wiley, New York (1976).}
\ref\KruJos{N. Joshi and M. D. Kruskal, in {\sl ``Painlev\'e Transcendents:
Their Asymptotics and Physical Applications"}, NATO ASI Series B: Vol. 278,
D. Levi and P. Winternitz Eds., Plenum Press, New York (1992).}
for more recent discussions) we make the change of variable
\eqn\VarCha{
u\ =\ t^{1/2}\,U\qquad;\qquad T\,=\,{4\over 5}\,t^{5/4}
}
in \Painleve, which becomes
\eqn\Ellip{
U''\,-\,6\,U^2\,+\,6\ =\ -\,{U'\over T}\,+\,{4\over 25}\,{U\over T^2}
}
As $|T|\to\infty$, $U$ is asymptotic to a Weirstrass elliptic $\wp$ function
$U_0(T,E_0)$, solution of
\eqn\Weirs{
(U_0')^2\ =\ 4\,U_0^3\,-\,12\,U_0\,+\,E_0
}
which is doubly periodic (with a lattice of double poles) with periods
\eqn\periods{
\Omega_{1,2}\,=\,\oint_{{\cal C}_{1,2}}\! dU_0\,
\left(4U_0^3-12U_0+E_0\right)^{-1/2}
}
where ${\cal C}_{1,2}$ are two contours encircling pairs of zeros of the
r.h.s. of \Weirs .
In a neighborhood of some $T=T_0$, one can treat $E_0$ as a slowly varying
variable $E_0(T)$. From \Ellip\ , in the local periods coordinates
$T-T_0=\Omega_1y^1+\Omega_2y^2$, $E_0$ varies as
\eqn\Eflow{
{\partial E_0\over\partial y^{1,2}}\ \simeq\,-\,{2\over T}\,{\cal J}_{1,2}(E_0)
\quad;\quad {\cal J}_{1,2}(E_0)\,=\,\oint_{{\cal C}_{1,2}}\!dU_0\,
\left(4U_0^3-12U_0+E_0\right)^{1/2}
}
Solving the flow equations \Eflow\ one can check that asymptotically, $E_0$
depends only on the argument of $T$, $\Theta={\rm Arg}(T)$, but not on
its modulus, and is solution of the two constraints
\eqn\constE{
{\rm Re}\left[{\rm e}^{i\Theta}\,{\cal J}_{1,2}(E_0)\right]\ =\ 0
}
But these constraints are exactly equivalent to \Constraint\ for \Fscal ,
once we identify $E_0=t^{-3/2}{\bf c}$ and use the fact that
$\Theta={5\over 4}{\rm Arg}(t)$.
\medskip

Before discussing the asymptotics for the loop operators, let us show how the
choice of weight contours $c_\pm$ in \ContComb\ fixes uniquely
the non-perturbative part of the solution of \Painleve.
Let us denote by $Z_N(\theta)$ (resp. $F_N(\theta)$ the partition function
\evInt\ (resp. it's Logarithm) for $N$ e.v.'s with the contour coefficients
\RealW .
Taking the derivative of $F$ w.r.t. $\theta$ singles out one of the e.v.'s
\eqn\thetder{
{dF_N \over d\theta}\ =\ i\, {N\over Z_N}\,{\bf C}_N\,\int_{{\cal C}_i}
d\lambda_0\,\int \prod_{i=1}^{N-1}\,d\lambda_i\,\Delta_{\scriptscriptstyle N}
(\lambda_i)^2\, {\rm e}^{-N\sum V(\lambda_i)}
}
where ${\cal C}_i$ is the contour going from ${\bar j}\infty$ to $j \infty$.
One estimates this integral by first integrating out the $N-1$ last e.v.'s
(by using the BIPZ method), in the effective potential
${\tilde V}(\lambda)= V(\lambda)+{1\over
N}(V(\lambda)-2\ln(\lambda_0-\lambda))$
modified by the first e.v.
The resulting effective potential for the first e.v. is in general complicated,
since it takes into account the back-reaction of this e.v. on the bulk
$N-1$ others e.v.'s. However it takes a simple form if $\lambda_0$ is close to
the end-point $\lambda_e$ of the e.v. distribution, since we obtain
\eqn\tdasym{
{dF_N \over d\theta}\ =\ {i\over 8\pi}\,\int_{{\cal C}_i} d\lambda_0\,
{1\over \lambda_0-\lambda_e}\,{\rm e}^{-N[\Gamma(\lambda_0)-\Gamma(\lambda_e)]}
\big(1+{\cal O}({1\over N})\big)}
where $\Gamma(\lambda)$ is the effective potential \effpot .
For the $m=2$ critical point, in the scaling regime $\Gamma'(p)=-2w(p)$, with
$w(p)$ given by \solpert . At large $N$ the integral \tdasym\ is dominated by
the instanton configuration of \Davpha , where the e.v. sits at the top
of the potential $p=t^{1/2}$. The result, including the contribution
of fluctuations around the saddle point, is
\eqn\npterms{
{dF\over d\theta}\ =\ -\,g_{\rm s}^{1/2}\,{3^{-3/4}\over 8\sqrt{\pi}}\,
t^{-5/8}\,{\rm e}^{-{1\over g_{\rm s}}\,{4\over 5}2\sqrt{3}\,t^{5/4}}\,
\big(1+{\cal O}({1\over N})\big)
}
where $g_{\rm s}=N^{-1}a^{-5/2}$ is the string coupling constant.
In the double scaling limit $g_{\rm s}=1$, and \npterms\ gives the
non-perturbative $\theta$-dependence of the string susceptibility $u=-F''$.
Using the fact that the triply truncated solutions $u_\pm$ correspond to
$\theta=\mp {i\over 2}$, one thus recovers the results of \SilYel \FokItsKit\
for the non-perturbative part of the simply truncated solutions of Painlev\'e
I.
\medskip

Finally, let us return to the construction of the loop operators.
It is very easy to check that with the generic choice of contours \ContComb,
the finite-$N$ loop equations of the matrix model are still satisfied
by the loop operators. It remains to understand what is the continuum limit
of these operators.
In string perturbation theory, the operator which creates a macroscopic loop
with momentum $p$ (conjugate to the loop length $\ell$),
$w(p)=\int_0^\infty d\ell\,{\rm e}^{-p\ell}\,w(\ell )$, can be expressed
in terms of matrix elements of the operator $Q$ given by \Qcont\ \BDSS .
For instance the one-loop correlator is
\eqn\oneloop{
\langle w(p)\rangle\ =\ \int_t^\infty dx\,\langle x|{1\over p+Q}|x\rangle
}
and the problem is to define the resolvent
$G(x;p)=\langle x|(p-Q)^{-1}|x\rangle$ for the generic real solutions of
\Painleve , with poles on the negative real axis. $G$ must
satisfy the Gelfand-Dikii equation
\eqn\GelDik{
-2\,G\,G''\,+{G'}^2\,+\,4\,(p+2u(x))\,G^2\ =\ 1
}
and generically $G$ has also double poles at the poles of $u$. In fact there is
a unique asymptotic prescription for $G(x)$ in the strong coupling regime
$x\to -\infty$ which is consistent with the contour prescription \ContComb\
(defined by the $\theta$-parameter), and the specific $u$.
If we perform the rescaling (similar to \VarCha )
\eqn\RescalG{
X\ =\ {4\over 5}\,x^{5/4}\quad;\quad
G\ =\ x^{-1/4}\,H\quad;\quad P\ =\ x^{-1/2}\,p
}
in the limit $|X|\to\infty$ \GelDik\ becomes
\eqn\GeDiAsy{
-2\,H\,H''\,+{H'}^2\,+\,4\,(P+2U_0(X))\,H^2\ =\ 1\,+\,{\cal O}\big({1\over X}
\big)
}
where $U_0(X)$ is the elliptic function given by \VarCha , which is solution of
\Weirs .
There is a unique solution of \GeDiAsy\ which is doubly periodic with the
same periods $\Omega_{1,2}(E_0)$ than $U_0$. It is given explicitly by
\eqn\solH{
H(X;P)\ =\ {P-U_0(X)\over\sqrt{4P^3-12P+E_0}}
}
This, together with \oneloop\ and the constraints \constE\ which fix $E_0$,
gives the same asymptotic expression for the one loop correlator $w(p)$
in the non-perturbative phase $t<0$ than the expression \Fscal\ that we have
obtained previously through the BIPZ approach.
This achieves the identification of our large $N$ non-perturbative solution
of the matrix model with the real solutions of the string equation \Painleve .
These loop operators will satisfy the loop equations \loopmoi  \DVV \FKN, at
variance with the operators constructed only in the perturbative phase with the
prescription of \DouShe \GroMig \BDSS .
Each loop operator will have a {\it single pole} in $t$ wherever the
string susceptibility has a double pole. This is in fact natural, since the
string susceptibility is the v.e.v. of two microscopic loops (puncture
operator).
\medskip

Let us summarize and discuss our results for the case of pure gravity.

The proposal of \FokItsKit\ to take real combination of complex integration
contours in the one matrix model to obtain real solutions of the Painlev\'e I
string equation for pure $2d$ gravity ($c=0$) has been formulated here in the
framework of the large $N$ solution of the matrix model \`a la BIPZ, i.e. in
terms of distribution of eigenvalues.
We have shown that the strong coupling phase, which corresponds to
negative values of the renormalized cosmological constant $t$, and in which the
string susceptibility has poles, can be described simply in term of splitting
of the mean-field distribution of the eigenvalues into two branches at the
end of the e.v. distribution.

These two different branches can be viewed as two different
topological sectors in the integral over the e.v.'s. and
the non-perturbative $\theta$-parameter which distinguishes the different
solutions of the string equation is simply the phase difference between these
two sectors, which has to be specified in the functional integral.
Therefore, at a formal level, each $\theta$ defines a $\theta$-vacuum of
$2d$ gravity, as in field theories with topological
sectors, such as $4d$ non-Abelian gauge theories or some $2d$ $\sigma$-models.
The non-perturbative effects associated to this $\theta$ parameter can be
estimated by simple instanton methods.

Finally, we have shown that, for each real solution of the string equation
($\theta$-vacuum), it is possible to construct in a consistent way
observables (loop operators), in such a way that the loop equations
should be satisfied non-perturbatively.
\medskip

Of course, many interesting questions are still open.

It is clear that one can define non-perturbatively the one matrix model for
general potential, and probably reconstruct by adequate choice of contours
the real non-perturbative solutions of the unstable even $m$ string equations.
We shall discuss below the case of the double well potential.
Similarly, the same recipe can be applied to the multi-matrix models, and
used to study the general $(p,q)$ string equations (although for the
multi matrix models there is no simple picture of the large $N$ limit in
term of e.v. distribution).

The fact that the loop equations are still valid non-perturbatively in the
framework discussed here is quite appealing. These equations can be
derived from various point of views: Dyson-Schwinger equations for the matrix
models, Virasoro constraints for the KdV hierarchy, recursion relations in
$2d$ topological gravity. This is at variance with other schemes which have
been proposed for defining non-perturbatively $2d$ gravity
\ref\MarPar{E. Marinari and G. Parisi, Phys. Lett. 247B (1990) 537.}
\ref\DaJoMo{S. Dalley, C. Johnson and T.R. Morris, Nucl. Phys. B 368
(1992) 625.}

One important issue has to be properly understood.
In the strong coupling phase ($t<0$) the partition function $Z=\exp (F)$
has zeros, and the loop operators have poles.
This implies that the non-perturbative real solutions of $2d$ gravity that we
have discussed here should suffer from non-perturbative violation of
positivity, even in the weak coupling regime $t>0$.
It remains to understand what this really means when one formulates these
solutions in term of string field theories in low dimensional backgrounds,
in particular for positivity and unitarity.

The $c=1$ matrix model solution studied in
\ref\cone{
E. Br\'ezin, V. A. Kazakov and Al. Zamolodchikov, Nucl. Phys. B 338 (1990) 673.
\hfill\break\noindent
D. Gross and N. Milkovi\'c, Phys. Lett. 238B (1990) 217.
\hfill\break\noindent
P. Ginsparg and J. Zinn-Justin, Phys. Lett. 240B (1990) 33.
}
does not suffer from the kind of instability of the $c=0$ model, since it
corresponds to free fermions in an inverted harmonic potential, with the
two wells of the potential filled at the same Fermi level.
Consequently, although the string perturbation theory for the $c=1$ model
is not Borel summable, there is a well defined summation prescription which
allows to reconstruct this non-perturbative solution.
Does the kind of ideas discussed here allow to construct other non-perturbative
solutions of the $c=1$ model?
\medskip

Finally let us briefly discuss the case of the double well potential
\eqn\dbwell{
V(\lambda)={1\over 2}\mu\lambda^2+{1\over 4}\lambda^4
}
For $\mu<0$ large enough, the e.v. are distributed along two cuts (symmetric
under $\lambda\leftrightarrow -\lambda$).
At the critical point $\mu_c$, the two cuts fuse (at $\lambda=0$)
into one segment. In the double scaling limit the string equation for this
critical point is the Painlev\'e \II\ equation
\ref\PII{M. R. Douglas, N. Seiberg and S. Shenker, Phys. Lett. 244B (1990) 381.
\hfill\break\noindent
V. Periwal and D. Shevitz, Phys. Rev. Lett. 64 (1990) 1326.}.
Recently, Brower et al. showed that by relaxing the parity condition
$\pi_n(\lambda)=(-1)^n\pi_n(-\lambda)$
on the orthonormal polynomial and on the associated solutions of the recurrence
equations, new symmetry breaking solutions of the model could be obtained
\BrDeJaTa .
This can be easily understood by considering the three independent paths of
integration for the potential \dbwell .
In addition to the real axis ${\cal C}_r$ we can also integrate over the
paths ${\cal C}_\pm$ going from $-\infty$ to $\pm i \infty$.
The most general weight factors for these paths which give a real partition
function are
\eqn\wdw{
c_r\,=\,1-2x\qquad;\qquad c_\pm\,=\,x\,\pm\,i\theta
}
If $c_\pm\ne 0$, in the strong coupling phase $\mu>\mu_c$ the e.v distribution
is no more the one cut solution but a cross-shaped distribution with
four cuts meeting at the origin.
Setting $x\ne 0$ breaks explicitly the symmetry
$\lambda\leftrightarrow -\lambda$
and should allow to recover the symmetry breaking solutions of \BrDeJaTa\
(which differ from the standard solution by subdominant terms of order
$1/N^2$).
Setting $x=0$ but $\theta\ne 0$ gives solutions which differ non-perturbatively
from the standard one, and which correspond to solutions of the Painlev\'e
\II\ equation with (simple) poles on the negative real axis.

These considerations can be extended easily to the multicut matrix models
studied in
\ref\CrnMoo{
\v C. Crnkovic and G. Moore, Phys. Lett. 257B (1991) 322.}.
\bigskip\noindent
{\bf Acknowledgments:}\par
I am very indebted to S. Jain for explaining to me the results of \BrDeJaTa\
prior to publication. I thank R. Conte, P. Di Francesco, J. Zinn-Justin and
J.-B. Zuber for very useful discussions, and P. Di Francesco for a careful
reading of the manuscript.

\listrefs
\listfigs
\bye